# Exact Quantum Query Algorithm for Error Detection Code Verification

Alina Vasilieva

Faculty of Computing, University of Latvia
Raina bulv. 29, LV-1459, Riga, Latvia
alina.vasilieva@gmail.com

**Abstract.** Quantum algorithms can be analyzed in a query model to compute Boolean functions. Function input is provided in a black box, and the aim is to compute the function value using as few queries to the black box as possible. A repetition code is an error detection scheme that repeats each bit of the original message *r* times. After a message with redundant bits is transmitted via a communication channel, it must be verified. If the received message consists of *r*-size blocks of equal bits, the conclusion is that there were no errors. The verification procedure can be interpreted as an application of a query algorithm, where input is a message to be checked. Classically, for *N*-bit message, values of all *N* variables must be queried. We present an exact quantum algorithm that uses only *N*/2 queries.

## 1 Introduction

Quantum computing is an exciting alternative way of computation, which is based on the laws of quantum mechanics. This branch of computer science is developing rapidly; various computational models exist, and this is a study of one of them.

Let $f(x_1, x_2, ..., x_N): \{0,1\}^N \rightarrow \{0,1\}$ be a Boolean function. We consider the black box model (also known as the query model), where a black box contains the input $X = (x_1, x_2, ..., x_N)$ and can be accessed by questioning $x_i$ values. The goal is to compute the value of the function. The complexity of a query algorithm is measured by the number of questions it asks. The classical version of this model is known as *decision trees* [1]. This computational model is widely applicable in software engineering. For instance, a database can be considered a black box, and, to speed up application performance, the goal is to reduce the number of database queries.

Quantum query algorithms can solve certain problems faster than classical algorithms. The best known and at the same time the simplest exact quantum algorithm for a total Boolean function was designed for the *XOR* function with *N*/2 questions vs. *N* questions required by classical algorithm [2]. The quantum query model differs from the quantum circuit model [2,3,4], and algorithm construction techniques for this model are less developed. The problem of quantum query algorithm construction is cardinally non-trivial. Although there are many lower bound and upper bound estimations of quantum query algorithm complexity [2,5,6,7], there are very few examples of original quantum query algorithms.

In this paper, we present a new exact quantum query algorithm for resolving a specific problem. The task is to verify a codeword message that has been encoded using *repetition code* for detecting errors [8] and has been transmitted across a communication channel. Considered repetition code simply duplicates each bit of the message. The verification procedure can be considered as an application of a query algorithm, where the codeword to be checked is contained in a black box. To verify the message in the classical way, we would need to access all bits. That is, for a codeword of length $N$, all $N$ queries to the black box would be required. We have developed an exact quantum query algorithm that requires only $N/2$ queries.

An exact algorithm always produces a correct answer with 100% probability. Another variation is to use a bounded-error model, where an error margin of 1/3 is allowed. It is well known that in the bounded-error model, a large difference between classical and quantum computation is possible. The complexity gap can be exponential as, for instance, in the Shor's algorithm case [9]. Another famous example is Grover's search algorithm that achieves a quadratic speed up [10]. However, in certain types of computer software, we cannot allow even a small probability of error, for example, in spacecraft, aircraft, or medical software. For this reason, the development of exact algorithms is extremely important.

Regarding exact quantum algorithms, the maximum speedup achieved as of now is half the number of queries compared with a classical deterministic case[1] [11]. The major open question is: is it possible to reduce the number of queries by more than 50%? In this paper, we present an algorithm that achieves a borderline gap of $N/2$ versus $N$.

## 2 Preliminaries

This section contains definitions and provides theoretical background on the subject. First, we introduce repetition codes for error detection and define a Boolean function for their verification. Then, we describe classical decision trees and show how to verify the simplest codeword in this model. We also show that, to verify an $N$-bit codeword classically, all $N$ queries are required. Next, we provide a brief overview of the basics of quantum computing. Finally, we describe the quantum black box model that is the subject of this paper.

### 2.1 Error Detection and Repetition Codes

In this article, we investigate a problem related to information transmission across a communication channel. The bit message is transmitted from a sender to a receiver. During that transfer, information may be corrupted. Because of the noise in a channel or adversary intervention some bits may disappear, or may be reverted, or even added. Various schemes exist to detect errors during transmission. In any case, a verification step is required after transmission. The received codeword is checked using defined rules and, as a result, a conclusion is made as to whether errors are present.

---

[1] Exact quantum algorithm with complexity $Q_E(f)<D(f)/2$ is not yet discovered for a total Boolean function. For partial Boolean functions this limitation can be exceeded. An excellent example is the Deutsch-Jozsa algorithm [12,13].

We consider a repetition error detection scheme known as repetition codes. A repetition code is a (r, n) coding scheme that repeats each n-bit block r times [8].

*Example:*
- Using a (3,1) repetition code, the message m=101 is encoded as c=111000111.
- Using a (2,2) repetition code, the message m=1011 is encoded as c=10101111.
- Using a (2,3) repetition code, m=111000 is encoded as c=111111000000.

Verification procedure for repetition code is rather simple – one just needs to check if in each group of r consecutive blocks of size n all blocks are equal.

In this article, we examine verification of the (2,1) repetition code. The verification process can be expressed naturally as a computing Boolean function in a query model. We assume that the codeword to be checked is located in a black box. We define the Boolean function to be computed by the query algorithm as follows:

**Definition 1.** *The Boolean function* $VERIFY_N(X)$, *where* $N = 2k$, $X = (x_1, x_2, ..., x_{2k})$ *is defined to have a value of "1" Iff variables are equal by pairs:*

$$VERIFY_{2k}(X) = \begin{cases} 1, \text{ if } x_1 = x_2 \text{ \& } x_3 = x_4 \text{ \& } x_5 = x_6 \text{ \&} ... \text{\& } x_{2k-1} = x_{2k} \\ 0, \text{ otherwise} \end{cases}$$

*Example:* The Boolean function $VERIFY_4(X)$ has the following accepting inputs:
$$\{0000, 0011, 1100, 1111\}.$$

The main result of this paper is an exact quantum query algorithm for $VERIFY_N(X)$, which calculates function value using N/2 quantum queries only.

## 2.2 Classical Decision Trees

The classical version of the query model is known as *decision trees* [1]. A black box contains the input $X = (x_1, x_2, ..., x_N)$ and can be accessed by questioning $x_i$ values. The algorithm must be able to determine the value of a function correctly for arbitrary input. The complexity of the algorithm is measured by the number of queries on the worst-case input. For more details, see the survey by Buhrman and de Wolf [1].

**Definition 2 [1].** *The **deterministic complexity** of a function f, denoted by **D(f)**, is the maximum number of questions that must be asked on any input by a deterministic algorithm for f.*

**Definition 3 [1].** *The **sensitivity** $s_x(f)$ **of f on input** $(x_1, x_2, ..., x_N)$ is the number of variables $x_i$ with the following property: $f(x_1, ..., x_i, ..., x_N) \neq f(x_1, ..., 1-x_i, ..., x_N)$. The **sensitivity of f** is $s(f) = \max_x s_x(f)$.*

It has been proved that $D(f) \geq s(f)$ [1].

Figure 1 demonstrates a classical deterministic decision tree, which computes $VERIFY_4(x_1, x_2, x_3, x_4)$. In this figure, circles represent queries, and rectangles represent output:

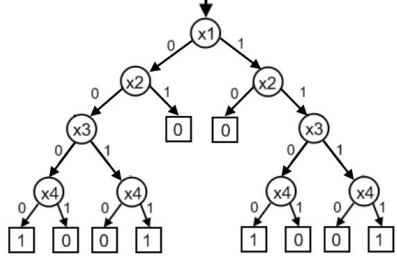

**Fig. 1**. Classical deterministic decision tree for computing $VERIFY_4(x_1, x_2, x_3, x_4)$.

**Theorem 1.** $D(VERIFY_N) = N$.

**Proof.** Check function sensitivity on any accepting input, for instance, on $X$=1111..11. Inversion of any bit will invert the function value, because a pair of bits with different values will appear. $s(VERIFY_N) = N \Rightarrow D(VERIFY_N) = N$.

### 2.3 Quantum Computing

This section briefly outlines the basic notions of quantum computing that are necessary to define the computational model used in this paper. For more details, see the textbooks by Nielsen and Chuang [3] and Kaye et al. [4].

An $n$-dimensional quantum pure state is a unit vector in a Hilbert space. Let $|0\rangle$, $|1\rangle$,..., $|n-1\rangle$ be an orthonormal basis for $\mathbb{C}^n$. Then, any state can be expressed as $|\psi\rangle = \sum_{i=0}^{n-1} \alpha_i |i\rangle$ for some $\alpha_i \in \mathbb{C}$. Since the norm of $|\psi\rangle$ is 1, we have $\sum_{i=0}^{n-1} |a_i|^2 = 1$. States $|0\rangle$, $|1\rangle$,..., $|n-1\rangle$ are called *basis states*. Any state of the form $\sum_{i=0}^{n-1} a_i |i\rangle$ is called a *superposition* of $|0\rangle,...,|n-1\rangle$. The coefficient $\alpha_i$ is called an *amplitude* of $|i\rangle$.

The state of a system can be changed by applying *unitary transformation*. The unitary transformation $U$ is a linear transformation on $\mathbb{C}^n$ that maps vectors of unit norm to vectors of unit norm. The *transpose* of a $m \times n$ matrix $A$ is the $n \times m$ matrix $A_{ij}^T = A_{ji}$ for $1 \leq i \leq n$, $1 \leq j \leq m$. We denote the *tensor product* of two matrices by $A \otimes B$.

The simplest case of quantum measurement is used in our model. It is the full measurement in the computation basis. Performing this measurement on a state $|\psi\rangle = \alpha_0 |0\rangle + ... + \alpha_{n-1} |n-1\rangle$ produces the outcome $i$ with a probability of $|\alpha_i|^2$. The measurement changes the state of the system to $|i\rangle$ and destroys the original state $|\psi\rangle$.

## 2.4 Quantum Query Model

The quantum query model is also known as the quantum black box model. This model is the quantum counterpart of decision trees and is intended for computing Boolean functions. For a detailed description, see the survey by Ambainis [6] and textbooks by Kaye, Laflamme, Mosca [4] and de Wolf [2].

A quantum computation with $T$ queries is a sequence of unitary transformations:
$$U_0 \to Q_0 \to U_1 \to Q_1 \to ... \to U_{T-1} \to Q_{T-1} \to U_T$$
$U_i$'s can be arbitrary unitary transformations that do not depend on input bits. $Q_i$'s are query transformations. Computation starts in the initial state $|\vec{0}\rangle$. Then we apply $U_0$, $Q_0,..., Q_{T-1}, U_T$ and measure the final state.

We use *ket* notation [3] to describe state vectors and algorithm flows:
$$|final\rangle = U_T \cdot Q_{T-1} \cdot ... \cdot Q_0 \cdot U_0 \cdot |\vec{0}\rangle.$$

We use the following definition of a query transformation: if input is a state $|\psi\rangle = \sum_i a_i |i\rangle$, then the output is $|\phi\rangle = \sum_i (-1)^{x_{k_i}} a_i |i\rangle$, where we can arbitrarily choose a variable assignment of $x_{k_i}$ for each basis state $|i\rangle$.

Formally, any transformation must be defined as a unitary matrix. The following is a matrix representation of a quantum black box query:

$$Q = \begin{pmatrix} (-1)^{X_{k_1}} & 0 & ... & 0 \\ 0 & (-1)^{X_{k_2}} & ... & 0 \\ ... & ... & ... & ... \\ 0 & 0 & ... & (-1)^{X_{k_m}} \end{pmatrix}$$

Each quantum basis state corresponds to the algorithm's output. We assign a value of a function to each output. The probability of obtaining the result $j \in \{0,1\}$ after executing an algorithm on input $X$ equals the sum of squared modulus of all amplitudes, which correspond to outputs with value $j$.

**Definition 4 [1].** *A quantum query algorithm **computes f exactly** if the output equals f(x) with a probability $p = 1$, for all $x \in \{0,1\}^N$. Complexity is denoted by $Q_E(f)$.*

Quantum query algorithms can be conveniently represented in diagrams, and we will use this approach in this paper.

## 3 Computing Function $VERIFY_N$ in a Quantum Query Model

In this section, we present the results of designing an exact quantum query algorithm for Boolean function $VERIFY_N(X)$. To define an algorithm, it is necessary to precisely describe all unitary transformations, all query matrices, and the measurement. We start from the simplest case of four variables and then show how to extend the algorithm to verify $N$-bit codewords. We have used a combinatorial approach to determine the structure of the algorithm, and have used *Mathematica*[©] [14] software developed by Wolfram Research to verify its correctness. In our approach, we have tried to employ the full power of quantum parallelism, also known as computing in a superposition.

### 3.1 Exact Quantum Query Algorithm for *VERIFY*$_4$

To familiarize the reader with the quantum query model and to build a base for extension, we demonstrate an algorithm for verification of 4-bit codewords. The algorithm flow is presented in Figure 2.

**Theorem 2.** *There exists an exact quantum query algorithm Q1 that computes the Boolean function VERIFY$_4$(X) using two queries:* $Q_E(Q1) = 2$.

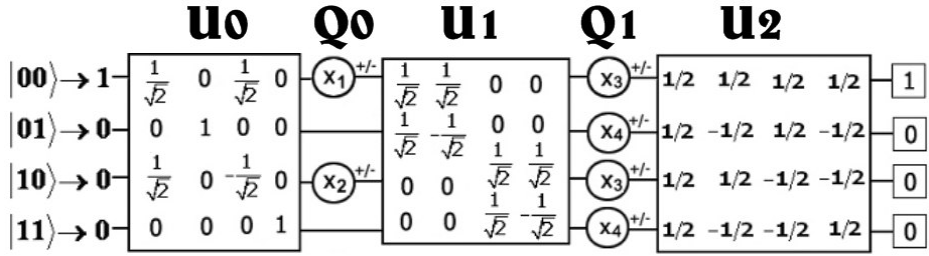

**Fig. 2**. Exact quantum query algorithm Q1 for computing the Boolean function *VERIFY$_4$*

The algorithm for computing the Boolean function *VERIFY$_4$* uses a 2-qubit quantum system. Each horizontal line corresponds to the amplitude of the basis state. Computation starts with amplitude distribution $|START\rangle = (1, 0, 0, 0)^T$. Three large rectangles correspond to the $4 \times 4$ unitary matrices $U_0$, $U_1$ and $U_2$. Two vertical layers of circles specify the queried variable order for queries $Q_0$ and $Q_1$. Finally, four small squares at the end of each horizontal line define the assigned function value for each basis state.

We demonstrate an example of computational flow for accepting input $X=1100$[2]:

$$|final\rangle = U_2Q_1U_1Q_0U_0|\vec{0}\rangle = U_2Q_1U_1Q_0U_0(1,0,0,0)^T =$$

$$= U_2Q_1U_1Q_0\left(\frac{1}{\sqrt{2}},0,\frac{1}{\sqrt{2}},0\right)^T = U_2Q_1U_1\left(-\frac{1}{\sqrt{2}},0,-\frac{1}{\sqrt{2}},0\right)^T =$$

$$= U_2Q_1\left(-\frac{1}{2},-\frac{1}{2},-\frac{1}{2},-\frac{1}{2}\right)^T = U_2\left(-\frac{1}{2},-\frac{1}{2},-\frac{1}{2},-\frac{1}{2}\right)^T =$$

$$= (-1,0,0,0)^T \xrightarrow{[Measure]} [\, f(1100) = 1 \text{ with a probability of } p = 1 \,]$$

---

[2] See the Appendix for results of computations for all inputs.

## 3.2 Exact Quantum Query Algorithm for VERIFY$_N$

This section contains the main result of this paper: a generalized algorithm for computing the Boolean function VERIFY$_N$. In the previous section, we demonstrated in detail the first algorithm in the sequence. Now, we will show how to extend our approach to verify codewords of length $N$.

**Theorem 3.** *The Boolean function VERIFY$_N(X)$ can be computed by an exact quantum query algorithm using N/2 queries:* $Q_E(\text{VERIFY}_N) = N/2$.

First of all, let us show how the main characteristics of the algorithm will change depending on the number of function variables.

**Table 1**. Main algorithm characteristics depending on number of variables.

| Variables | Qubits | Amplitudes | Queries |
|---|---|---|---|
| 4 | 2 | 4 | 2 |
| 6 | 3 | 8 | 3 |
| … | … | … | … |
| $N$ | $N/2$ | $2^{N/2}$ | $N/2$ |

The main task is to define algorithm flow, i.e., the sequence of transformations. Unitary transformations and query matrices must be constructed in such a way to produce the correct result on any possible input. We are considering exact quantum algorithms, and this adds an additional condition to the final amplitude distribution that must be obtained.

We introduce an algorithm that will construct all required transformation matrices for a specified $N$. Then obtained transformations must be applied to the initial state in a specified order. Finally, the measurement must be performed on the final amplitude distribution.

The algorithm is described in Table 2. The algorithm was implemented using *Mathematica*© software[3], and its correctness was verified by a computer program.

**Table 2**. Exact quantum query algorithm for computing the Boolean function VERIFY$_N$.

| 1. Setup |
|---|
| Boolean function to be computed: $\text{VERIFY}_N(x_1, x_2, ..., x_N)$ |
| Number of queries: $T = N/2$ |
| Number of qubits: $T$ |
| Number of amplitudes (dimension of Hilbert space): $K = 2^T = 2^{N/2}$ |
| Algorithm flow: $\vert START \rangle \to U_1 \to Q_1 \to ... \to U_T \to Q_T \to U_{FINAL} \to [Measure]$ |

---
[3] See the *Mathematica* program code in the Appendix.

## 2. Algorithm structure construction

```
FOR (i=1 to T){
    STEP 1: Calculate a set of indices:
```

$$|IND| = 2^i \, ; \, IND = \left\{ind_1, ind_2, ..., ind_{2^i}\right\} \, ;$$

$$IND = \left\{ j \cdot \frac{K}{2^i} + 1 \, \middle| \, j \in \{0, 1, .., (2^i - 1)\} \right\}$$

```
    STEP 2: Construct matrices U_i and Q_i:
        Initialize U_i with the identity matrix U_i = I_K
        Initialize Q_i with the identity matrix Q_i = I_K
        index=1;
        WHILE(index< 2^i ){
            t1=IND[index]    // index^th element from the set IND
            t2=IND[index+1]

            Replace elements of U_i and Q_i :
```

$$U_{t_1,t_1} = U_{t_1,t_2} = U_{t_2,t_1} = \frac{1}{\sqrt{2}}$$

$$U_{t_2,t_2} = -\frac{1}{\sqrt{2}}$$

$$Q_{t_1,t_1} = (-1)^{X_{2i-1}}$$

$$Q_{t_2,t_2} = (-1)^{X_{2i}}$$

```
            index = index + 2;
        }
}
```

STEP 3: Final transformation – $U_{FINAL} = H^{\otimes T}$, where $H = \frac{1}{\sqrt{2}}\begin{pmatrix} 1 & 1 \\ 1 & -1 \end{pmatrix}$.

STEP 4: Initial state – $|START\rangle = (1, 0, 0, ..., 0)^T$

STEP 5: Measurement – the only accepting state is $|\vec{0}\rangle = |000..0\rangle$.

## 3. Algorithm application

Execute the algorithm on input $X$ by applying a constructed unitary and query transformation in the following order:

$$|START\rangle \rightarrow U_1 \rightarrow Q_1 \rightarrow ... \rightarrow U_T \rightarrow Q_T \rightarrow U_{FINAL} \rightarrow [Measure]$$

### 3.3 Algorithm Illustration for $VERIFY_6$

To make the described algorithm more transparent and to avoid possible misunderstandings, we illustrate it for the case of six variables.

*Setup:*

Boolean function to be computed: $VERIFY_6(x_1, x_2, ..., x_6)$
Number of queries and number of qubits = 3.
Number of amplitudes (dimension of Hilbert space) = 8.
Algorithm flow:
$$|START\rangle \to U_1 \to Q_1 \to U_2 \to Q_2 \to U_3 \to Q_3 \to H^{\otimes 3} \to [Measure]$$

*Algorithm structure construction:*

$i=1$: $IND = \{1, 5\}$

$$U_1 = \begin{pmatrix} \frac{1}{\sqrt{2}} & 0 & 0 & 0 & \frac{1}{\sqrt{2}} & 0 & 0 & 0 \\ 0 & 1 & 0 & 0 & 0 & 0 & 0 & 0 \\ 0 & 0 & 1 & 0 & 0 & 0 & 0 & 0 \\ 0 & 0 & 0 & 1 & 0 & 0 & 0 & 0 \\ \frac{1}{\sqrt{2}} & 0 & 0 & 0 & -\frac{1}{\sqrt{2}} & 0 & 0 & 0 \\ 0 & 0 & 0 & 0 & 0 & 1 & 0 & 0 \\ 0 & 0 & 0 & 0 & 0 & 0 & 1 & 0 \\ 0 & 0 & 0 & 0 & 0 & 0 & 0 & 1 \end{pmatrix} \quad Q_1 = \begin{pmatrix} (-1)^{X1} & 0 & 0 & 0 & 0 & 0 & 0 & 0 \\ 0 & 1 & 0 & 0 & 0 & 0 & 0 & 0 \\ 0 & 0 & 1 & 0 & 0 & 0 & 0 & 0 \\ 0 & 0 & 0 & 1 & 0 & 0 & 0 & 0 \\ 0 & 0 & 0 & 0 & (-1)^{X2} & 0 & 0 & 0 \\ 0 & 0 & 0 & 0 & 0 & 1 & 0 & 0 \\ 0 & 0 & 0 & 0 & 0 & 0 & 1 & 0 \\ 0 & 0 & 0 & 0 & 0 & 0 & 0 & 1 \end{pmatrix}$$

$i=2$: $IND = \{1, 3, 5, 7\}$

$$U_2 = \begin{pmatrix} \frac{1}{\sqrt{2}} & 0 & \frac{1}{\sqrt{2}} & 0 & 0 & 0 & 0 & 0 \\ 0 & 1 & 0 & 0 & 0 & 0 & 0 & 0 \\ \frac{1}{\sqrt{2}} & 0 & -\frac{1}{\sqrt{2}} & 0 & 0 & 0 & 0 & 0 \\ 0 & 0 & 0 & 1 & 0 & 0 & 0 & 0 \\ 0 & 0 & 0 & 0 & \frac{1}{\sqrt{2}} & 0 & \frac{1}{\sqrt{2}} & 0 \\ 0 & 0 & 0 & 0 & 0 & 1 & 0 & 0 \\ 0 & 0 & 0 & 0 & \frac{1}{\sqrt{2}} & 0 & -\frac{1}{\sqrt{2}} & 0 \\ 0 & 0 & 0 & 0 & 0 & 0 & 0 & 1 \end{pmatrix} \quad Q_2 = \begin{pmatrix} (-1)^{X3} & 0 & 0 & 0 & 0 & 0 & 0 & 0 \\ 0 & 1 & 0 & 0 & 0 & 0 & 0 & 0 \\ 0 & 0 & (-1)^{X4} & 0 & 0 & 0 & 0 & 0 \\ 0 & 0 & 0 & 1 & 0 & 0 & 0 & 0 \\ 0 & 0 & 0 & 0 & (-1)^{X3} & 0 & 0 & 0 \\ 0 & 0 & 0 & 0 & 0 & 1 & 0 & 0 \\ 0 & 0 & 0 & 0 & 0 & 0 & (-1)^{X4} & 0 \\ 0 & 0 & 0 & 0 & 0 & 0 & 0 & 1 \end{pmatrix}$$

$i=2$: $IND = \{1, 2, 3, 4, 5, 6, 7\}$

$$U_3 = \begin{pmatrix} \frac{1}{\sqrt{2}} & \frac{1}{\sqrt{2}} & 0 & 0 & 0 & 0 & 0 & 0 \\ \frac{1}{\sqrt{2}} & -\frac{1}{\sqrt{2}} & 0 & 0 & 0 & 0 & 0 & 0 \\ 0 & 0 & \frac{1}{\sqrt{2}} & \frac{1}{\sqrt{2}} & 0 & 0 & 0 & 0 \\ 0 & 0 & \frac{1}{\sqrt{2}} & -\frac{1}{\sqrt{2}} & 0 & 0 & 0 & 0 \\ 0 & 0 & 0 & 0 & \frac{1}{\sqrt{2}} & \frac{1}{\sqrt{2}} & 0 & 0 \\ 0 & 0 & 0 & 0 & \frac{1}{\sqrt{2}} & -\frac{1}{\sqrt{2}} & 0 & 0 \\ 0 & 0 & 0 & 0 & 0 & 0 & \frac{1}{\sqrt{2}} & \frac{1}{\sqrt{2}} \\ 0 & 0 & 0 & 0 & 0 & 0 & \frac{1}{\sqrt{2}} & -\frac{1}{\sqrt{2}} \end{pmatrix} \quad Q_3 = \begin{pmatrix} (-1)^{X5} & 0 & 0 & 0 & 0 & 0 & 0 & 0 \\ 0 & (-1)^{X6} & 0 & 0 & 0 & 0 & 0 & 0 \\ 0 & 0 & (-1)^{X5} & 0 & 0 & 0 & 0 & 0 \\ 0 & 0 & 0 & (-1)^{X6} & 0 & 0 & 0 & 0 \\ 0 & 0 & 0 & 0 & (-1)^{X5} & 0 & 0 & 0 \\ 0 & 0 & 0 & 0 & 0 & (-1)^{X6} & 0 & 0 \\ 0 & 0 & 0 & 0 & 0 & 0 & (-1)^{X5} & 0 \\ 0 & 0 & 0 & 0 & 0 & 0 & 0 & (-1)^{X6} \end{pmatrix}$$

$$U_{final} = \begin{pmatrix} \frac{1}{2\sqrt{2}} & \frac{1}{2\sqrt{2}} & \frac{1}{2\sqrt{2}} & \frac{1}{2\sqrt{2}} & \frac{1}{2\sqrt{2}} & \frac{1}{2\sqrt{2}} & \frac{1}{2\sqrt{2}} & \frac{1}{2\sqrt{2}} \\ \frac{1}{2\sqrt{2}} & -\frac{1}{2\sqrt{2}} & \frac{1}{2\sqrt{2}} & -\frac{1}{2\sqrt{2}} & \frac{1}{2\sqrt{2}} & -\frac{1}{2\sqrt{2}} & \frac{1}{2\sqrt{2}} & -\frac{1}{2\sqrt{2}} \\ \frac{1}{2\sqrt{2}} & \frac{1}{2\sqrt{2}} & -\frac{1}{2\sqrt{2}} & -\frac{1}{2\sqrt{2}} & \frac{1}{2\sqrt{2}} & \frac{1}{2\sqrt{2}} & -\frac{1}{2\sqrt{2}} & -\frac{1}{2\sqrt{2}} \\ \frac{1}{2\sqrt{2}} & -\frac{1}{2\sqrt{2}} & -\frac{1}{2\sqrt{2}} & \frac{1}{2\sqrt{2}} & \frac{1}{2\sqrt{2}} & -\frac{1}{2\sqrt{2}} & -\frac{1}{2\sqrt{2}} & \frac{1}{2\sqrt{2}} \\ \frac{1}{2\sqrt{2}} & \frac{1}{2\sqrt{2}} & \frac{1}{2\sqrt{2}} & \frac{1}{2\sqrt{2}} & -\frac{1}{2\sqrt{2}} & -\frac{1}{2\sqrt{2}} & -\frac{1}{2\sqrt{2}} & -\frac{1}{2\sqrt{2}} \\ \frac{1}{2\sqrt{2}} & -\frac{1}{2\sqrt{2}} & \frac{1}{2\sqrt{2}} & -\frac{1}{2\sqrt{2}} & -\frac{1}{2\sqrt{2}} & \frac{1}{2\sqrt{2}} & -\frac{1}{2\sqrt{2}} & \frac{1}{2\sqrt{2}} \\ \frac{1}{2\sqrt{2}} & \frac{1}{2\sqrt{2}} & -\frac{1}{2\sqrt{2}} & -\frac{1}{2\sqrt{2}} & -\frac{1}{2\sqrt{2}} & -\frac{1}{2\sqrt{2}} & \frac{1}{2\sqrt{2}} & \frac{1}{2\sqrt{2}} \\ \frac{1}{2\sqrt{2}} & -\frac{1}{2\sqrt{2}} & -\frac{1}{2\sqrt{2}} & \frac{1}{2\sqrt{2}} & -\frac{1}{2\sqrt{2}} & \frac{1}{2\sqrt{2}} & \frac{1}{2\sqrt{2}} & -\frac{1}{2\sqrt{2}} \end{pmatrix}$$

### 3.4 Algorithm Analysis

To improve intuition and understanding, general algorithm for verification of *N*-bit codeword can be visualized as an abstract tree (see Figure 3). We start at the top (Level 1) with state vector that has exactly one amplitude initialized to $\alpha = 1$.

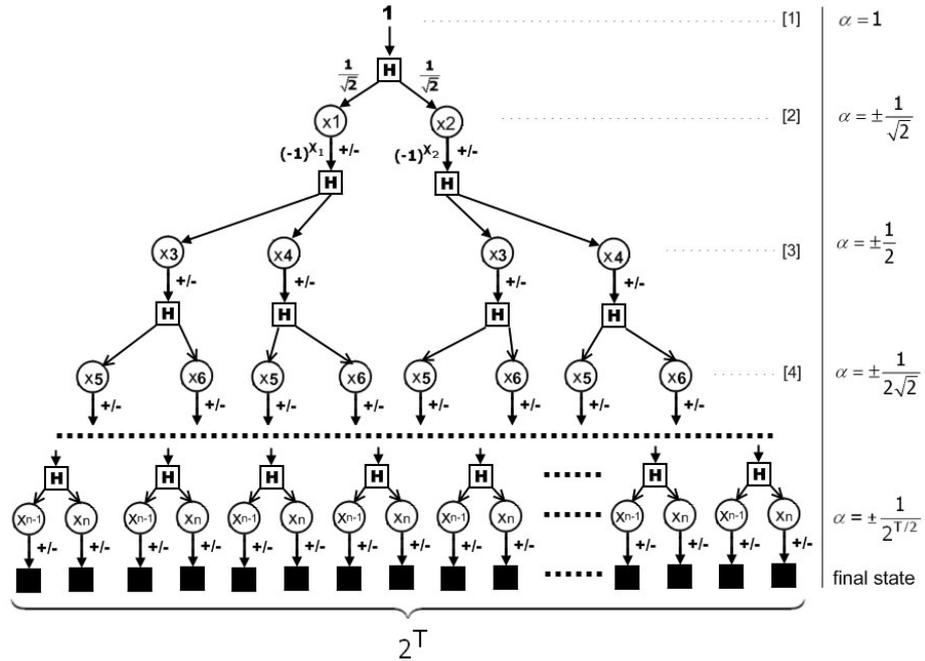

**Fig. 3** Visualization of the quantum query algorithm as an abstract tree.

Then Hadamard transform splits this amplitude between two basis states, so on the Level 2 we have state vector with two amplitudes set to $\alpha = \frac{1}{\sqrt{2}}$. Then the first query follows, which may change sign of amplitudes to the opposite depending on the values of queried variables. Next, Hadamard transforms nested in $U_2$ split amplitudes once again, and on the Level 3 there is a state vector with four amplitudes set to $\alpha = \pm\frac{1}{2}$ each. Next query followed by amplitude split – and on the Level 4 there are eight amplitudes in a state vector set to $\alpha = \pm\frac{1}{2\sqrt{2}}$, and so on. On the last bottom level, after $T$ queries, we have state vector in a form $\sum_{i=0}^{2^T-1} \alpha_i |i\rangle$, where $\alpha_i = \pm\frac{1}{2^{T/2}}$.

Queries and unitary transformations are formed and combined in such a way, that if values of function variables are equal by pairs ($x_1 = x_2$ & $x_3 = x_4$ and so on), then in the final state vector signs of all amplitudes will be identical. At the same time, the first row of matrix $U_{FINAL} = H^{\otimes T}$ consists of equal elements $+\frac{1}{2^{T/2}}$. It means that application of $U_{FINAL}$ will join together all amplitudes and results in the state vector with $\alpha = 1$ in the first position. So, the measurement will output the state $|\vec{0}\rangle = |00...0\rangle$ with 100% probability. This is the accepting state → $VERIFY_N(X) = 1$.

If algorithm is executed on rejecting input, i.e., there is at least one pair of variables with different values, then after all $T$ queries number of $+\frac{1}{2^{T/2}}$ and $-\frac{1}{2^{T/2}}$ amplitudes in state vector will be equal. This is provided by the algorithm structure. After multiplication with $U_{FINAL}$ the value of the first amplitude will be zero, so there is no probability to obtain $|\vec{0}\rangle$ state after the measurement → $VERIFY_N(X) = 0$.

Examined cases show that our algorithm always computes Boolean function $VERIFY_N(X)$ with probability of 1.

## 4 Application for a String Equality Problem

Our quantum algorithm can be adapted for solving such computational problem as testing if two binary strings are equal. This is a well-known task, which can be used as a subroutine in various algorithms.

Our quantum algorithm for Boolean function $VERIFY_N$ checks whether variables are equal by pairs, i.e., $x_1 = x_2$ & $x_3 = x_4$ & ... & $x_{N-1} = x_N$. On the other hand, we can consider that our algorithm is checking whether two binary strings, $Y = x_1 x_3 x_5 ... x_{N-1}$ and $Z = x_2 x_4 x_6 ... x_N$, are equal. Therefore, our algorithm can be easily used not only to verify repetition codes, but also for checking the equality of binary strings.

## 5  Conclusion

In this research paper, we investigated the verification of error detection codes. We have represented the verification procedure as an application of a query algorithm to an input codeword contained in a black box. We have presented an exact quantum query algorithm, which allows verifying a codeword of length $N$ using only $N/2$ queries to the black box. Our algorithm saves exactly half the number of queries comparing to the classical case. This result repeats the largest difference between classical and quantum algorithm complexity for a total Boolean function known today in this model. We believe that our algorithm is an important and useful addition to the collection of exact quantum query algorithms.

We see many possibilities for future research in the area of quantum query algorithm design. The most significant open question still remains: is it possible to increase algorithm performance more than two times using quantum tools? We believe it may be possible. Next, there are many computational tasks waiting for efficient solution in a quantum setting. Regarding the verification of repetition codes, we would like to be able to verify not only (2,1) code, but also an arbitrary ($r,n$) code. Another fundamental goal is to develop a framework for building efficient ad-hoc quantum query algorithms for arbitrary Boolean functions.

## References

1. H. Buhrman and R. de Wolf: Complexity Measures and Decision Tree Complexity: A Survey. Theoretical Computer Science, v. 288(1): 21–43 (2002).
2. R. de Wolf: Quantum Computing and Communication Complexity. University of Amsterdam (2001).
3. M. Nielsen, I. Chuang: Quantum Computation and Quantum Information. Cambridge University Press (2000).
4. P.Kaye, R.Laflamme, M.Mosca: An Introduction to Quantum Computing. Oxford (2007).
5. A.Ambainis: Quantum query algorithms and lower bounds (survey article). In Proceedings of FOTFS III, Trends on Logic, vol. 23 (2004), pp. 15-32.
6. A.Ambainis and R. de Wolf: Average-case quantum query complexity. Journal of Physics A 34, pp. 6741–6754 (2001).
7. A.Ambainis: Polynomial degree vs. quantum query complexity. Journal of Computer and System Sciences 72, pp. 220–238 (2006).
8. T. M. Cover and J. A. Thomas: Elements of Information Theory. pp. 209—212, Wiley-Interscience, (1991).
9. P. W. Shor: Polynomial time algorithms for prime factorization and discrete logarithms on a quantum computer. SIAM Journal on Computing, 26(5):1484-1509 (1997).
10. L. Grover: A fast quantum mechanical algorithm for database search. In Proceedings of 28[th] STOC'96, pp. 212. –219 (1996).
11. A.Ambainis. Personal communication, April 2009.
12. D. Deutsch and R. Jozsa: Rapid solutions of problems by quantum computation. In Proceedings of the Royal Society of London, volume A 439, pp. 553-558 (1992).
13. R. Cleve, A. Ekert, C. Macchiavello, and M. Mosca: Quantum algorithms revisited. In Proceedings of the Royal Society of London, volume A 454, pp. 339–354 (1998).
14. Wolfram Research, *Mathematica*[©], http://www.wolfram.com/

# 6 Appendix

## 6.1 Exact Quantum Query Algorithm for *VERIFY₄*

In Section 3.1 we presented an exact quantum query algorithm for computing the Boolean function *VERIFY*$_4$ using two quantum queries. Table below provides details about computation process for each input.

**Table 3.** Quantum query algorithm computation process for *VERIFY*$_4$.

| X | after $Q_0U_0\|\vec{0}\rangle$ | after $Q_1U_1Q_0U_0\|\vec{0}\rangle$ | final state | result |
|---|---|---|---|---|
| 0000 | $\left(\frac{1}{\sqrt{2}},0,\frac{1}{\sqrt{2}},0\right)$ | $\left(\frac{1}{2},\frac{1}{2},\frac{1}{2},\frac{1}{2}\right)$ | (1,0,0,0) | **1** |
| 0001 | $\left(\frac{1}{\sqrt{2}},0,\frac{1}{\sqrt{2}},0\right)$ | $\left(\frac{1}{2},-\frac{1}{2},\frac{1}{2},-\frac{1}{2}\right)$ | (0,1,0,0) | **0** |
| 0010 | $\left(\frac{1}{\sqrt{2}},0,\frac{1}{\sqrt{2}},0\right)$ | $\left(-\frac{1}{2},\frac{1}{2},-\frac{1}{2},\frac{1}{2}\right)$ | (0,-1,0,0) | **0** |
| 0011 | $\left(\frac{1}{\sqrt{2}},0,\frac{1}{\sqrt{2}},0\right)$ | $\left(-\frac{1}{2},-\frac{1}{2},-\frac{1}{2},-\frac{1}{2}\right)$ | (-1,0,0,0) | **1** |
| 0100 | $\left(\frac{1}{\sqrt{2}},0,-\frac{1}{\sqrt{2}},0\right)$ | $\left(\frac{1}{2},\frac{1}{2},-\frac{1}{2},-\frac{1}{2}\right)$ | (0,0,1,0) | **0** |
| 0101 | $\left(\frac{1}{\sqrt{2}},0,-\frac{1}{\sqrt{2}},0\right)$ | $\left(\frac{1}{2},-\frac{1}{2},-\frac{1}{2},\frac{1}{2}\right)$ | (0,0,0,1) | **0** |
| 0110 | $\left(\frac{1}{\sqrt{2}},0,-\frac{1}{\sqrt{2}},0\right)$ | $\left(-\frac{1}{2},\frac{1}{2},\frac{1}{2},-\frac{1}{2}\right)$ | (0,0,0,-1) | **0** |
| 0111 | $\left(\frac{1}{\sqrt{2}},0,-\frac{1}{\sqrt{2}},0\right)$ | $\left(-\frac{1}{2},-\frac{1}{2},\frac{1}{2},\frac{1}{2}\right)$ | (0,0,-1,0) | **0** |
| 1000 | $\left(-\frac{1}{\sqrt{2}},0,\frac{1}{\sqrt{2}},0\right)$ | $\left(-\frac{1}{2},-\frac{1}{2},\frac{1}{2},\frac{1}{2}\right)$ | (0,0,-1,0) | **0** |
| 1001 | $\left(-\frac{1}{\sqrt{2}},0,\frac{1}{\sqrt{2}},0\right)$ | $\left(-\frac{1}{2},\frac{1}{2},\frac{1}{2},-\frac{1}{2}\right)$ | (0,0,0,-1) | **0** |
| 1010 | $\left(-\frac{1}{\sqrt{2}},0,\frac{1}{\sqrt{2}},0\right)$ | $\left(\frac{1}{2},-\frac{1}{2},-\frac{1}{2},\frac{1}{2}\right)$ | (0,0,0,1) | **0** |

| 1011 | $\left(-\dfrac{1}{\sqrt{2}}, 0, \dfrac{1}{\sqrt{2}}, 0\right)$ | $\left(\dfrac{1}{2}, \dfrac{1}{2}, -\dfrac{1}{2}, -\dfrac{1}{2}\right)$ | (0,0,1,0) | **0** |
|---|---|---|---|---|
| 1100 | $\left(-\dfrac{1}{\sqrt{2}}, 0, -\dfrac{1}{\sqrt{2}}, 0\right)$ | $\left(-\dfrac{1}{2}, -\dfrac{1}{2}, -\dfrac{1}{2}, -\dfrac{1}{2}\right)$ | (-1,0,0,0) | **1** |
| 1101 | $\left(-\dfrac{1}{\sqrt{2}}, 0, -\dfrac{1}{\sqrt{2}}, 0\right)$ | $\left(-\dfrac{1}{2}, \dfrac{1}{2}, -\dfrac{1}{2}, \dfrac{1}{2}\right)$ | (0,-1,0,0) | **0** |
| 1110 | $\left(-\dfrac{1}{\sqrt{2}}, 0, -\dfrac{1}{\sqrt{2}}, 0\right)$ | $\left(\dfrac{1}{2}, -\dfrac{1}{2}, \dfrac{1}{2}, -\dfrac{1}{2}\right)$ | (0,1,0,0) | **0** |
| 1111 | $\left(-\dfrac{1}{\sqrt{2}}, 0, -\dfrac{1}{\sqrt{2}}, 0\right)$ | $\left(\dfrac{1}{2}, \dfrac{1}{2}, \dfrac{1}{2}, \dfrac{1}{2}\right)$ | (1,0,0,0) | **1** |

## 6.2 *Mathematica*© Program Source Code for General Algorithm

```
n=4;       (* Initialization parameter – number of function variables *)

T=n/2;     (* Number of queries and qubits *)
K=2^T;     (* Number of amplitudes *)
Print["n=",n," T=",T," K=",K];

UnitaryTransformations={};    (* Real unitary transformation matrices *)
QueriesPrintable={};          (* Printable query matrices *)

(* CONSTRUCT UNITARY TRANSFORMATIONS AND PRINTABLE QUERY MATRICES *)

For[i=1,i<=T, i++,

  (* Step 1 – Calculate set of indices *)

  Indices={};
  For[j=0,j<2^i,j++,
    Indices=Append[Indices,j*K/(2^i)+1];
  ];
  Print[Indices];

  (* Step 2 – Construct Ui and Qi *)

  Ui=IdentityMatrix[K];
  Qi=IdentityMatrix[K];

  index=1;
  While[index<2^i,
    t1=Indices[[index]];
    t2=Indices[[index+1]];

    Ui=ReplacePart[Ui,1/Sqrt[2],{t1,t1}];
    Ui=ReplacePart[Ui,1/Sqrt[2],{t1,t2}];
    Ui=ReplacePart[Ui,1/Sqrt[2],{t2,t1}];
    Ui=ReplacePart[Ui,-1/Sqrt[2],{t2,t2}];
```

```
      Qi=ReplacePart[Qi,(-1)^StringJoin["X",ToString[2i-1]],{t1,t1}];
      Qi=ReplacePart[Qi,(-1)^StringJoin["X",ToString[2i]],{t2,t2}];

      index=index+2;
    ];
  UnitaryTransformations=Append[UnitaryTransformations,Ui];
  QueriesPrintable=Append[QueriesPrintable,Qi];
];

(* Construct final transformation *)

H={{1/Sqrt[2],1/Sqrt[2]},{1/Sqrt[2],-1/Sqrt[2]}};
UFinal=H;
For[i=1,i<T,i++,
    UFinal=BlockMatrix[Outer[Times,UFinal,H]]
];

START={{1}};
MEASUREMENT={1};
For[i=1,i<K, i++,
    START=Append[START,{0}];
    MEASUREMENT=Append[MEASUREMENT,0];
];

(* Print all constructed matrices *)

For[i=1,i<=Length[UnitaryTransformations], i++,
    Print["U",i,"=",UnitaryTransformations[[i]]//MatrixForm];
    Print["Q",i,"=",QueriesPrintable[[i]]//MatrixForm];
];
Print["U final=",UFinal//MatrixForm];
Print["START=",START];
Print["MEASUREMENT=",MEASUREMENT];

(* START ALGORITHM EXECUTION FOR EACH INPUT VECTOR *)

For[inputAsNumber=0,inputAsNumber<2^n, inputAsNumber++,
    Binary={};
    num=inputAsNumber;
    For[j=0,j<n, j++,
      result=Floor[num/2];
      Binary=Append[Binary,num-result*2];
      num=result;
    ];
    X=Reverse[Binary];

    (* Calculate real query matrices *)
    QueryTransformations={};
    For[i=1,i<=T, i++,
      Indices={};
      For[j=0,j<2^i,j++,
        Indices=Append[Indices,j*K/(2^i)+1];
      ];

      index=1;
      Qi=IdentityMatrix[K];
      While[index<2^i,
        t1=Indices[[index]];
```

```
            t2=Indices[[index+1]];
            Qi=ReplacePart[Qi,(-1)^X[[2i-1]],{t1,t1}];
            Qi=ReplacePart[Qi,(-1)^X[[2i]],{t2,t2}];
            index=index+2;
            ];
        QueryTransformations=Append[QueryTransformations,Qi];
    ];

    (* EXECUTE ALGORITHM FLOW *)

    RESULT=START;
    For[i=1,i<=Length[UnitaryTransformations], i++,
    RESULT=QueryTransformations[[i]].UnitaryTransformations[[i]].RESULT;
    ];
    RESULT=UFinal.RESULT;
    F=MEASUREMENT.RESULT;
    F=Abs[F[[1]]];

    Print[""];
    Print["X=",X];
    Print["Final distribution: ",RESULT];
    Print["Function value: ", F];
];
```